\documentclass{PoS}

\usepackage{epsfig,multicol,multirow}
\usepackage{amsmath,subfigure,latexsym,amssymb}

\def\lsi{\raise0.3ex\hbox{$<$\kern-0.75em\raise-1.1ex\hbox{$\sim$}}}
\def\gsi{\raise0.3ex\hbox{$>$\kern-0.75em\raise-1.1ex\hbox{$\sim$}}}
\def\backder{\raise1.4ex\hbox{$\leftarrow$\kern-0.75em\raise-1.4ex\hbox{$\parti
al$}}}
\def\forder{\raise1.4ex\hbox{$\rightarrow$\kern-0.75em\raise-1.4ex\hbox{$\parti
al$}}}

\newcommand{\be}{\begin{equation}}
\newcommand{\ee}{\end{equation}}
\newcommand{\nn}{\nonumber}
\newcommand{\bea}{\begin{eqnarray}}
\newcommand{\eea}{\end{eqnarray}}

\newcommand{\R}{{\kern+.25em\sf{R}\kern-.78em\sf{I} \kern+.78em\kern-.25em}}
\newcommand{\RR}{{\kern+.25em\sf{R}\kern-.6em\sf{I} \kern+.6em\kern-.25em}}
\newcommand{\N}{{\kern+.25em\sf{N}\kern-.78em\sf{I} \kern+.78em\kern-.25em}}
\newcommand{\C}{{\kern+.25em\sf{C}\kern-.50em\sf{I} \kern+.50em\kern-.25em}}

\title{Topological Summation of Observables Measured with Dynamical Overlap Fermions}

\ShortTitle{Topological Summation with Dynamical Overlap Fermions}

\author{\speaker{Wolfgang Bietenholz} \\%
        \  \\
        John von Neumann Institut f\"{u}r Computing NIC \\
        Deutsches Elektron-Synchrotron DESY \\
        Platanenallee 6 \\
        15738 Zeuthen, Germany \\
        E-mail: \email{bietenho@ifh.de}}

\author{Ivan Hip\\
        \  \\
        Faculty of Geotechnical Engineering \\
        University of Zagreb \\
        Hallerova aleja 7 \\
        42000 Vara\v{z}din, Croatia \\
        E-mail: \email{ivan.hip@gmail.com \\ }}

\abstract{HMC histories for light dynamical overlap fermions 
tend to stay in a fixed topological sector for many trajectories, 
so that the different sectors are not sampled properly.
Therefore the suitable summation of observables, 
which have been measured in separate sectors, is a major challenge.
We explore several techniques for this issue,
based on data for the chiral condensate and the (analogue of the)
pion mass in the 2-flavour Schwinger model with
dynamical overlap-hypercube fermions.}

\FullConference{The XXVI International Symposium on Lattice Field Theory \\
                 July 14 - 19, 2008\\
                 Williamsburg, Virginia, USA}

\begin{document}

\section{The 2-flavour Schwinger model}

The Schwinger model (QED$_{2}$) on a Euclidean plane
is characterised by the Lagrangian
\be
{\cal L}( \bar \Psi , \Psi , A_{\mu}) = \bar \Psi (x) \, [
\gamma_{\mu} ( {\rm i} \partial_{\mu} + g A_{\mu}) + m ] \,
\Psi (x) + \frac{1}{2} F_{\mu \nu}(x) F_{\mu \nu}(x) \ .
\ee
Analytic results are obtained in the bosonised form, {\it e.g.}\
for the chiral condensate 
$\Sigma = - \langle \bar \Psi \Psi \rangle$ and the
``meson'' masses (iso-triplet and iso-singlet),
with two degenerate flavours of mass $ m \ll g\,$,
\be  \label{cond}
\Sigma (m) 
= 0.388 \dots (m g^{2})^{1/3} ~~~ \cite{Smilga} \ , ~~~
M_{\pi} = 2.008 \dots (m^{2} g)^{1/3} ~~~ 
\cite{Smilga} \ , ~~~
M_{\eta} = \sqrt{ \frac{2 g^{2}}{\pi} + M_{\pi}^{2} }
~~~ \cite{Ivan} \ .
\ee


Here we consider a lattice formulation with compact link variables
$U_{\mu,x} \in U(1)$ and the plaquette gauge action.
For the fermions we use the overlap-hypercube Dirac operator 
\cite{WBEPJC,WBIH}
\be
D_{\rm ovHF} = ( 1 - \frac{m}{2}) D_{\rm ovHF}^{(0)} + m \ , \quad
D_{\rm ovHF}^{(0)} = 1 + (D_{\rm HF} -1) / \sqrt{
(D_{\rm HF}^{\dagger} -1) (D_{\rm HF} -1) } \ . \label{overlap}
\ee
$D_{\rm HF}$ is a truncated perfect hypercube fermion operator, which 
is $\gamma_{5}$-Hermitian, 
$D_{\rm HF}^{\dagger} = \gamma_{5} D_{\rm HF} \gamma_{5}\,$.
Due to the use of the overlap formula \cite{Neu} in eq.\ (\ref{overlap}),
$D_{\rm ovHF}^{(0)}$ solves the Ginsparg-Wilson relation in its
simplest form \cite{Has} , $\{ D_{\rm ovHF}^{(0)}, \gamma_{5} \} = 
2 D_{\rm ovHF}^{(0)} \gamma_{5} D_{\rm ovHF}^{(0)}$, which implies
a lattice modified, exact chiral symmetry \cite{ML}.
It reproduces the axial anomaly correctly in all topological 
sectors \cite{DAWB}.

In the free case $D_{\rm HF}$ is discussed in Refs.\ \cite{WBEPJC,WBIH}
(we refer to the version denoted as CO-HF). 
It is approximately chiral already, hence it changes only little 
in the transition to the overlap operator,
\be  \label{simi}
D_{\rm HF} \approx D_{\rm ovHF} \ ,
\ee 
in contrast to the Wilson kernel, which is used in
the standard overlap operator \cite{Neu}. 
Further virtues of $D_{\rm HF}$, which are based on the
renormalisation group construction, like improved scaling and 
approximate rotation symmetry are essentially inherited by
$D_{\rm ovHF}$. Also long-range couplings are only turned on
slightly, again due to the similarity (\ref{simi}), which strongly
improves the degree of locality compared to the standard overlap 
operator. As a related
property, the condition number of the operator in the inverse
square root of eq.\ (\ref{overlap}) is strongly reduced.

The form of $D_{\rm ovHF}[U]$ interacting through $U(1)$ gauge fields
was introduced in Ref.\ \cite{WBIH}. All the above virtues
were tested and confirmed extensively for the 2-flavour
Schwinger model. In that case, the configurations were generated
quenched, but their contributions to the measurements were
re-weighted with the fermion determinant \cite{WBIH,Q2}. 
Also in quenched QCD,
a drastically improved locality and approximate rotation symmetry
have been confirmed for $D_{\rm ovHF}[U]$ \
\cite{QCD}.

\section{Simulation with dynamical overlap fermions}

Hybrid Monte Carlo (HMC) simulations with chiral fermions are
tedious and still at an early stage.
However, in addition to the virtues listed above, the property (\ref{simi})
also facilitates HMC simulations: a low polynomial in $D_{\rm HF}$
can be used for the HMC force and for short trajectories
a useful acceptance rate persists \cite{Lat06,prep} (in the extreme 
case of using directly $D_{\rm HF}$, however, its volume dependence 
was considered unsatisfactory \cite{Quatsch}).

We performed HMC simulations with dynamical overlap-hypercube fermions
at $\beta = 1/g^{2} = 5$ on $L \times L$ lattices
with two fermion flavours of mass $m$, where $L$ and $m$ take
the following values:
\bea
{\underline{L=16}} & : & m = 0.01 \ , \ 0.03 \ , \ 0.06 \ , \ 0.09 \ , \
0.12 \ , \ 0.18 \ , \ 0.24 \ . \nn \\
{\underline{m=0.01}} & : & L = 16 \ , \ 20 \ , \ 24 \ , \ 28 \ , \ 32 \ . \nn
\eea
We did not study a continuum extrapolation, but the
lattices are fine (plaquette values $\approx 0.9$), so we rely on 
small lattice artifacts (moreover $O(a)$ scaling artifacts are ruled
out since we are using Ginsparg-Wilson fermions).

Our algorithm fulfils the conceptual conditions like detailed balance
and area conservation \cite{Lat06}. 
The essential practical question is
the acceptance rate in the Metropolis step at the end of the
HMC trajectories (it employs $D_{\rm ovHF}$ to precision $10^{-16}$).
For $L=16$ with trajectory length $\ell = 1/8$
we obtained acceptance rates in the range $0.3 \dots 0.5$, with 
only a mild dependence on the fermion mass \cite{Lat06}.
In Figure \ref{accL} we show the dependence on the lattice size
$L$ at very light mass $m=0.01$. The trajectory length
is decreased for $L= 16,\ 20, \ 24, \ 28, \ 32$ to
$\ell = 0.125, \ 0.0625, \ 0.05, \ 0.04, \ 0.03$, respectively.
This keeps the acceptance rates in the same range, and
large statistics allows for a selection of well decorrelated
configurations.
\FIGURE{
  \centering
\includegraphics[angle=270,width=.5\linewidth]{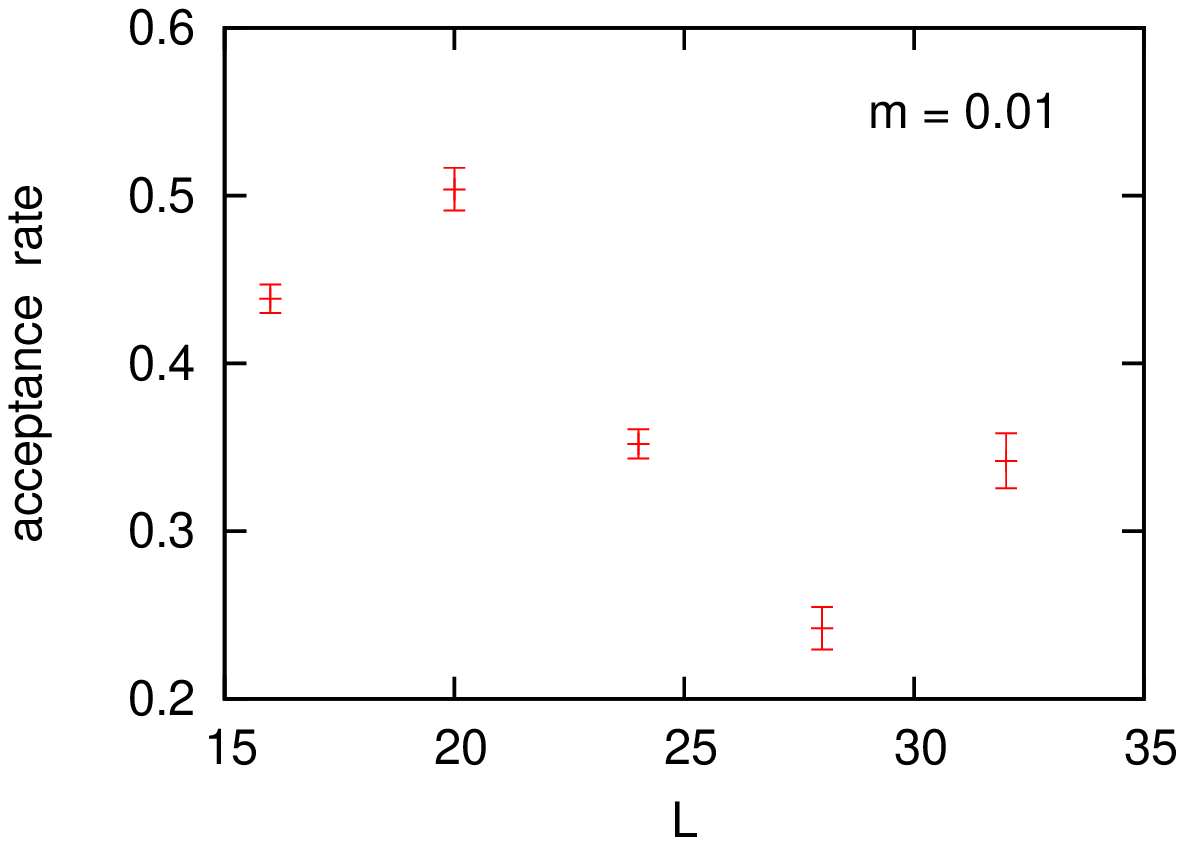}
\caption{The acceptance rate of the HMC trajectories
at $m=0.01$ as a function of the lattice size $L$. Note that
the trajectory length is reduced for increasing volume.}
\label{accL}
}
A major challenge for the simulation with light dynamical overlap 
fermions --- in addition to the huge computation time request in QCD
--- is that the histories at weak gauge coupling
perform only very few topological transitions (which require
a temporary deformation to rough configurations). 
We refer to the clean definition, which identifies the topological charge 
with the fermionic index $\nu$ \cite{Has}. In all our considerations
only the absolute value $| \nu |$ matters.

\section{The chiral condensate $\Sigma$}

For a configuration in volume $V = L^{2}$,
the chiral condensate is given by
\be
\Sigma = \frac{1}{V} \sum_{i} \frac{1}{| \lambda_{i}|+m} :=
\frac{|\nu |}{mV} + \varepsilon_{| \nu |} \ .
\ee
$\lambda_{i}$ are the Dirac eigenvalues mapped stereographically
onto the imaginary axis,
$\lambda_{i} \to \lambda_{i} /(1 - \lambda_{i}/2)$, 
and the sum runs over all of them. The last term defines the quantity  
$\varepsilon_{| \nu |}\,$.

In Ref.\ \cite{Lat07} we discussed the determination of $\Sigma$
based on the lowest non-zero eigenvalues and Random Matrix Theory 
(RMT). Since $\Sigma (m=0) =0$, this is not the 
setting
that RMT usually refers to, so we are probing 
{\it terra incognita}. 
In fact we observed microscopic eigenvalue densities, which are not 
described by any known RMT formula. Nevertheless, the RMT interpretation
of the ratio $\langle \lambda_{1} \rangle_{\nu =0}
/ \langle \lambda_{1} \rangle_{|\nu |=1}$
yields results for $\Sigma$, which agree well with the 
predicted value (\ref{cond}) over a broad parameter range.
($\langle \dots \rangle_{| \nu |}$ denotes an expectation
value restricted to configurations of charge $| \nu |$.)

\begin{figure}[h!]
\vspace*{-3mm}
\hspace*{-3mm} 
\includegraphics[angle=270,width=.54\linewidth]{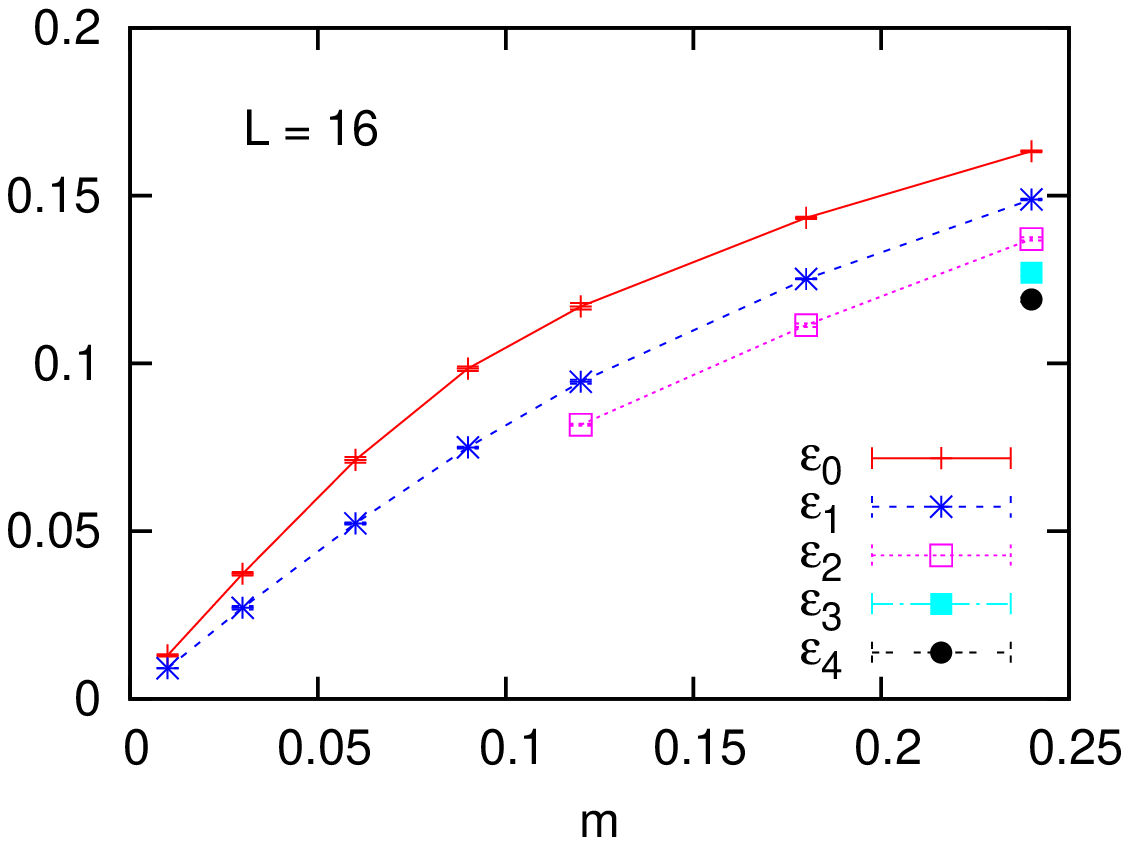}
\hspace*{-7mm}  
\includegraphics[angle=270,width=.54\linewidth]{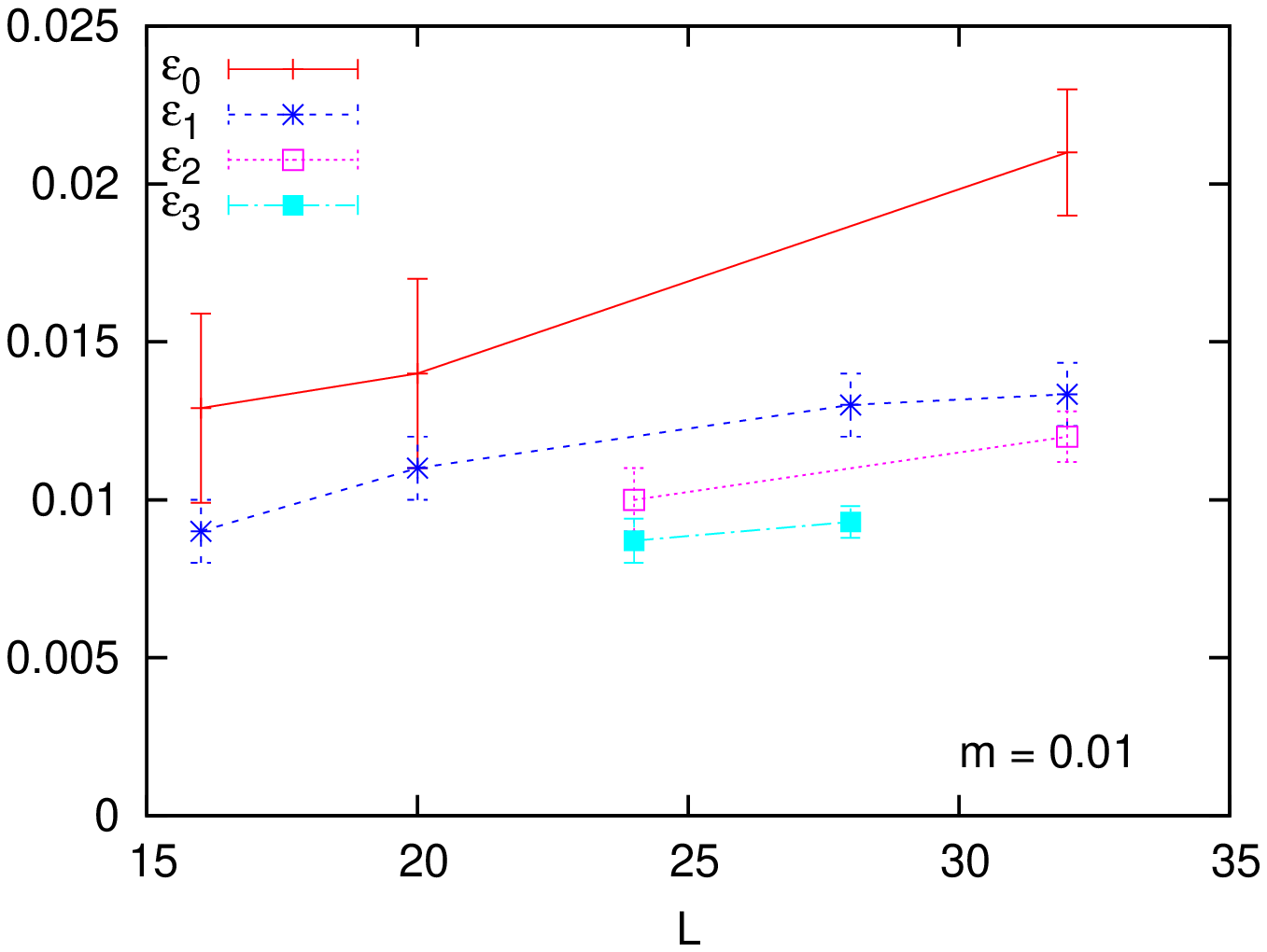}
\caption{The contributions of the non-zero modes to the chiral
condensate, $\varepsilon_{| \nu |}$, measured in the topological
sectors $\pm \nu $. We show the mass dependence at
$L=16$ (left) and the lattice size dependence at $m=0.01$ (right).}
\label{Sigmafig}
\end{figure}

Here we address the {\em direct} measurement of $\Sigma$ based on the 
full Dirac spectrum. Our results are shown in Figure \ref{Sigmafig}.
At small masses $m$, the zero mode contribution
strongly dominates for topologically charged configurations,
so that the term $\varepsilon_{| \nu |}$ is a minor correction.
As a generic property of stochastic Hermitian operators
(such as $\gamma_{5} D_{\rm ovHF}$), zero modes repel 
low lying non-zero modes, which implies the hierarchy
\be  \label{ugl1}
\varepsilon_{0} > \varepsilon_{1} > \varepsilon_{2} \dots
\ee
at fixed $m$ and $L$. This is observed consistently in
Figure \ref{Sigmafig} (on the left). As we increase the volume
at fixed $m$, more eigenvalues are accumulated near zero, and we infer
\be \label{ugl2}
\varepsilon_{i} (V_{1}) > \varepsilon_{i} (V_{2}) \qquad
{\rm for} \quad V_{1} > V_{2} \ .
\ee
Also that property is observed consistently from our data,
as Figure \ref{Sigmafig} (on the right) shows.

\subsection{Method 1 : Gaussian summation}

\begin{figure}[h!]
\vspace*{-2.5mm}
\hspace*{-5mm} 
\includegraphics[angle=270,width=.55\linewidth]{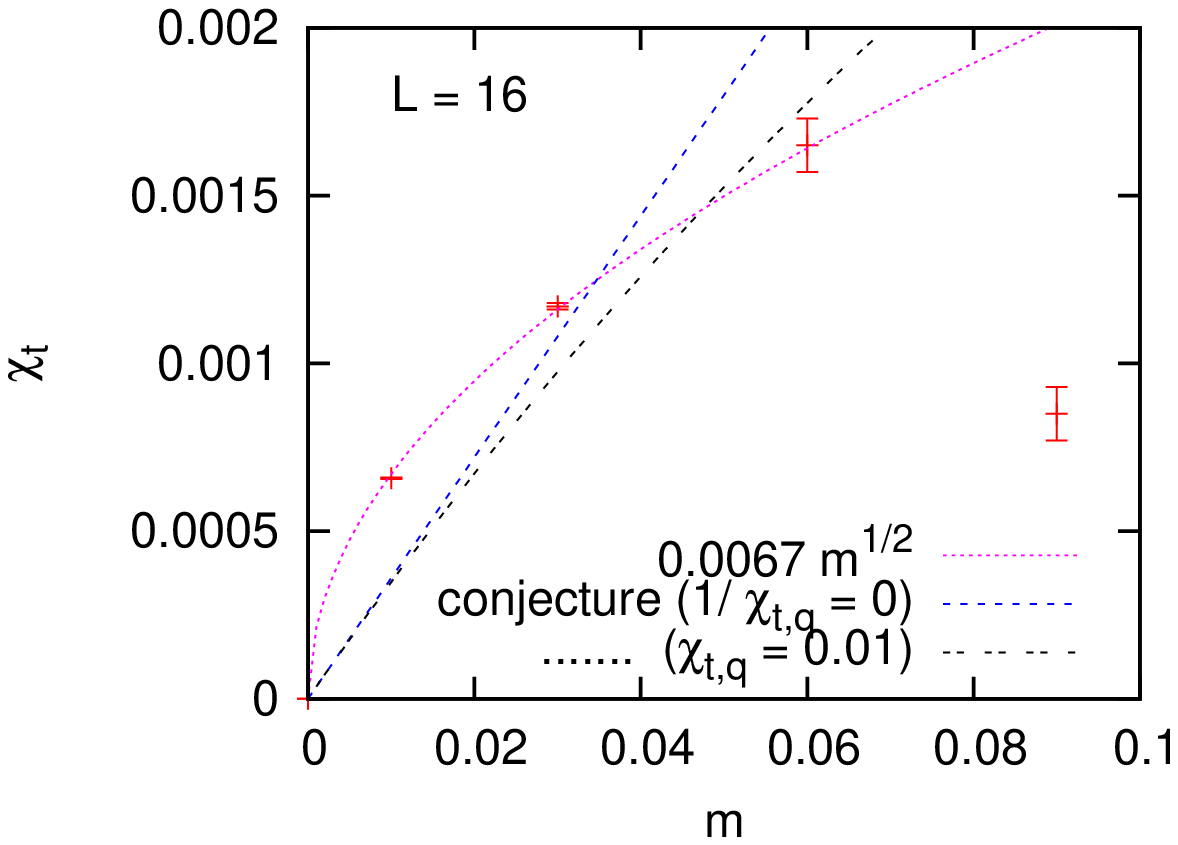}
\hspace*{-5mm}  
\includegraphics[angle=270,width=.55\linewidth]{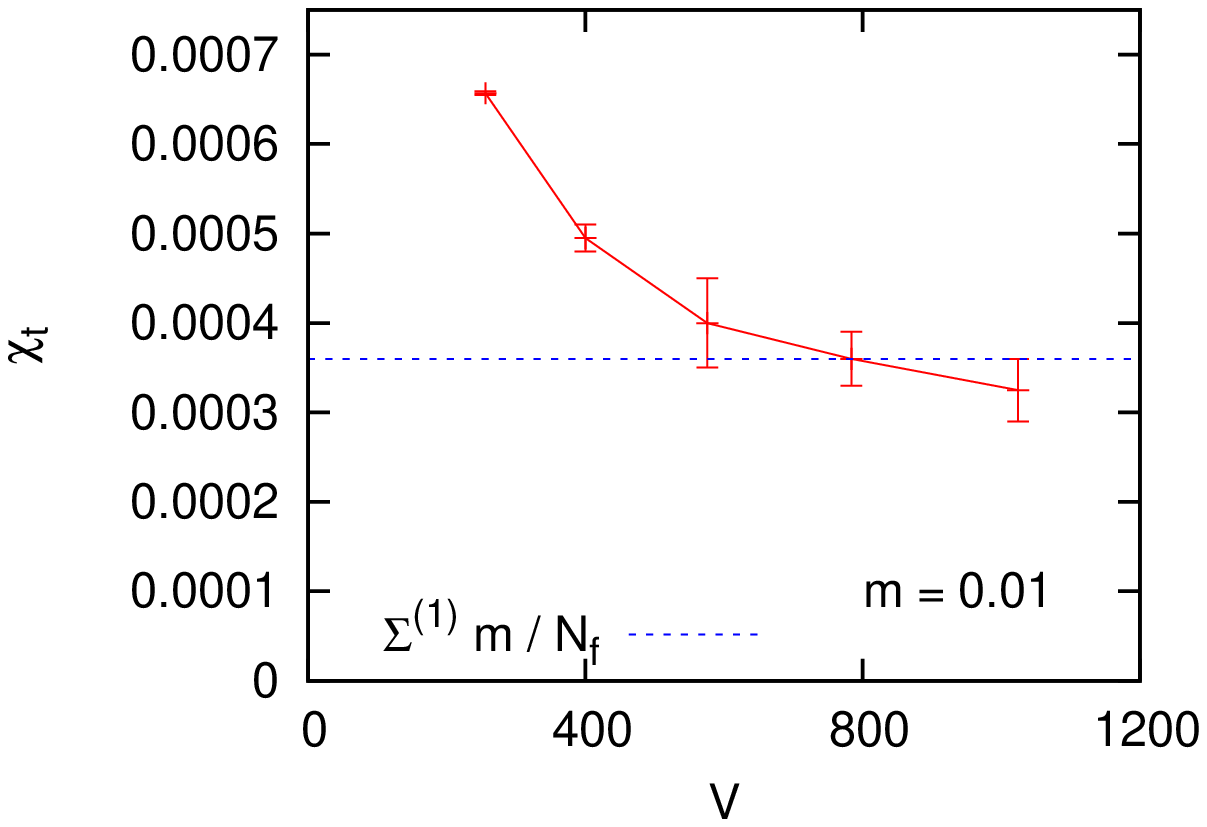}
\caption{Results for the topological susceptibility $\chi_{t}$
for light fermions, obtained by assuming a Gaussian distribution
of the topological charges and matching the theoretical value of $\Sigma$.
The plot on the left suggests a behaviour $\chi_{t} \propto \sqrt{m}$
at small masses. 
The plot on the right shows a nice convergence
to the value conjectured in Ref.\ \cite{Durr} (in infinite volume),
from where we also adapted the quenched susceptibility $\chi_{t,q}$.}
\label{chifig}
\end{figure}
Our first method to sum over the topologies
relies on the assumption that the probability
distribution of the charges $\nu$ is Gaussian.
Indeed general experience shows
that possible deviations from a Gaussian distribution tend to 
be small. Here our assumption implies
\bea
\Sigma &=& \sum_{\nu = -\infty}^{\infty} p(|\nu|) \ \Sigma_{|\nu|} \ ,
\quad {\rm where} \quad p(|\nu|) = \frac{\exp \{ - \nu^{2} / (2 V \chi_{t}) \}}
{\sum_{\nu} \exp \{ - \nu^{2} / (2 V \chi_{t}) \}} \ , \nn \\
\Sigma_{|\nu|} &:=& - \langle \bar \psi \psi \rangle_{| \nu |}
\quad {\rm and} \quad \chi_{t} = \langle \nu^{2} \rangle / V
\quad {\rm is~the~topological~susceptibility.}  \label{Gaussdis}
\eea
Assume we have measured results for $\Sigma_{0} \dots \Sigma_{Q}$.
This constrains the values at all higher charges based on
eq.\ (\ref{ugl1}),
\be
\frac{| \nu |}{mV} < \Sigma_{|\nu |} < \frac{| \nu |}{mV} + 
\varepsilon_{Q} \ .
\ee
For $L=24$ and $28$ we also have gaps at low $| \nu |$,
where we insert the obvious bounds based on eq.\ (\ref{ugl2}).
Thus we obtain --- with remarkably mild uncertainties --- the full
set of $\Sigma_{|\nu |}$ to be inserted in eq.\ (\ref{Gaussdis}).
The only unknown parameter is the susceptibility $\chi_{t}$,
which we {\em tune} so that the sum reproduces the theoretical
$\Sigma$ value (\ref{cond}). Thus we obtain the results for $\chi_{t}$
shown in Figure \ref{chifig}.
For very light fermions they suggest $\chi_{t} \propto \sqrt{m}$ 
(in a fixed volume), see plot on the left.

Alternative results (with quenched configurations and re-weighting)
were given in Ref.\ \cite{DuHo}. On the theoretical side, Ref.\ \cite{Durr}
conjectured for $N_{f}$ degenerate flavours in the large volume limit
\be  \label{chidu}
\frac{1}{\chi_{t}} = \frac{N_{f}}{\Sigma^{(1)}m} + 
\frac{1}{\chi_{t,q}} \ , \qquad
\Sigma^{(1)} := \Sigma_{N_{f}=1}(m=0) \simeq 0.16 \, g \ ,
\ee
where 
$\chi_{t,q} = \chi_{t} (m \to \infty)$ is the quenched value.
As we increase the volume at $m=0.01$, our results
converge to the vicinity of this prediction,
see Figure \ref{chifig} on the right.
\FIGURE{
  \centering
\includegraphics[angle=270,width=.54\linewidth]{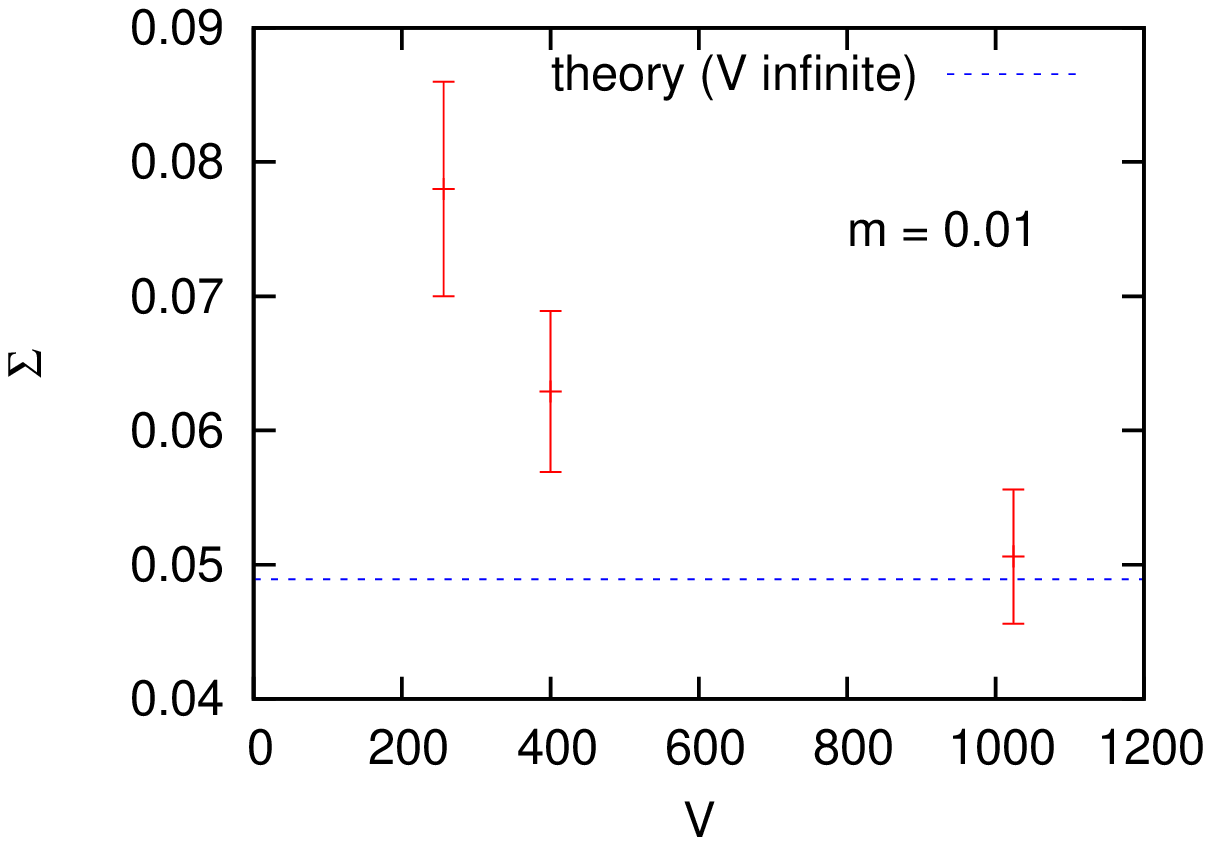}
\caption{The chiral condensate $\Sigma$ at $m=0.01$ as a function
of the volume, obtained from measurements of various $\Sigma_{|\nu |}$
values and the approximate summation formula (\ref{sumfo}).}
\label{Sigmafig2}
}

\subsection{Method 2 : Summation formula}

The first method used the theoretically predicted $\Sigma$ value
as an input to extract $\chi_{t}$. Now we try to evaluate
$\Sigma$ itself from our data. We apply an approximation formula 
\cite{prep}, in exact analogy to a formula for the pion mass 
derived in Ref.\ \cite{BCNW},
\be  \label{sumfo}
\Sigma_{|\nu |} \approx \Sigma - \frac{A}{V} + \nu^{2} 
\frac{B}{V^{2}} \ , \quad A = \frac{\alpha}{\chi_{t}}
\ , \quad B = \frac{\alpha}{\chi_{t}^{2}} \ .
\ee
The derivation involves a Fourier transform between
$\nu$ and the vacuum angle $\theta$, which is treated by 
the stationary phase approximation.\\
Formula (\ref{sumfo}) involves three unknown parameters. Two of
them, $\Sigma$ and $\chi_{t} = A/B$ are of physical interest
(unlike $\alpha$). We capture them if we manage to determine $A$ 
and $B$. \\
$\bullet$ At fixed $m$ and $V$ we can extract $B$, {\it e.g.}\
from $\Sigma_{0}$ and $\Sigma_{1}$. \\
$\bullet$ At fixed $m$ in two volumes $V_{1} \neq V_{2}$, we can
evaluate $A$, {\it e.g.}\ from $\Sigma_{0}(V_{1})$ and 
$\Sigma_{0}(V_{2})$. \\
\hspace*{2.5mm} $\Sigma_{|\nu |}$ values in three volumes
determine all three parameters.

For a combined approach along these lines \cite{prep}, 
we arrived at the results
for $\Sigma$ shown in Figure \ref{Sigmafig2}. As in Figure \ref{chifig} 
(on the right) we see for increasing volume a flow towards the value
(\ref{cond}), which was predicted theoretically (in infinite volume).

\subsection{Method 3 : Density correlation}

A third method has been suggested in Refs.\ \cite{Jap1}.
The derivation is similar to Ref.\ \cite{BCNW},\footnote{In 
both cases the approximation should be best for small
$| \nu | \,$, so that $\langle \nu^{2} \rangle \gg | \nu |$.}
and it leads to a formula for finite size effects in the correlation
of the topological charge density $\rho\,$,
\be
\langle \rho (x) \rho (y) \rangle_{| \nu |} \to
\frac{1}{V} \Big( \frac{\nu^{2}}{V} - \chi_{t} - 
\frac{c_{4}}{2 \chi_{t} V} \Big) + O(V^{-3}) \ . 
\qquad \qquad \qquad \qquad \qquad \qquad 
\ee
(The kurtosis $c_{4} = (3 \langle \nu^{2} \rangle^{2} - 
\langle \nu^{4} \rangle)/V$ is a measure for the deviation from
a Gaussian charge distribution.) The formula refers to the 
asymptotic behaviour at large $|x - y|$. This requires
a large volume, and unfortunately this also means that the
signal to extract $\chi_{t}$ is likely to be too small for a
conclusive measurement. Later Ref.\ \cite{Jap2} considered
instead the $\eta'$-correlator of the pseudo-scalar density, 
which obeys the analogous formula, and where a sensible signal 
was found, even in QCD with dynamical overlap quark.
Also in our project this method is under investigation 
\cite{prep}.

\vspace{-1mm}
\section{Meson masses}
\vspace{-1mm}

\begin{figure}
\begin{center}
\hspace*{-2mm}
\includegraphics[angle=270,width=.5\linewidth]{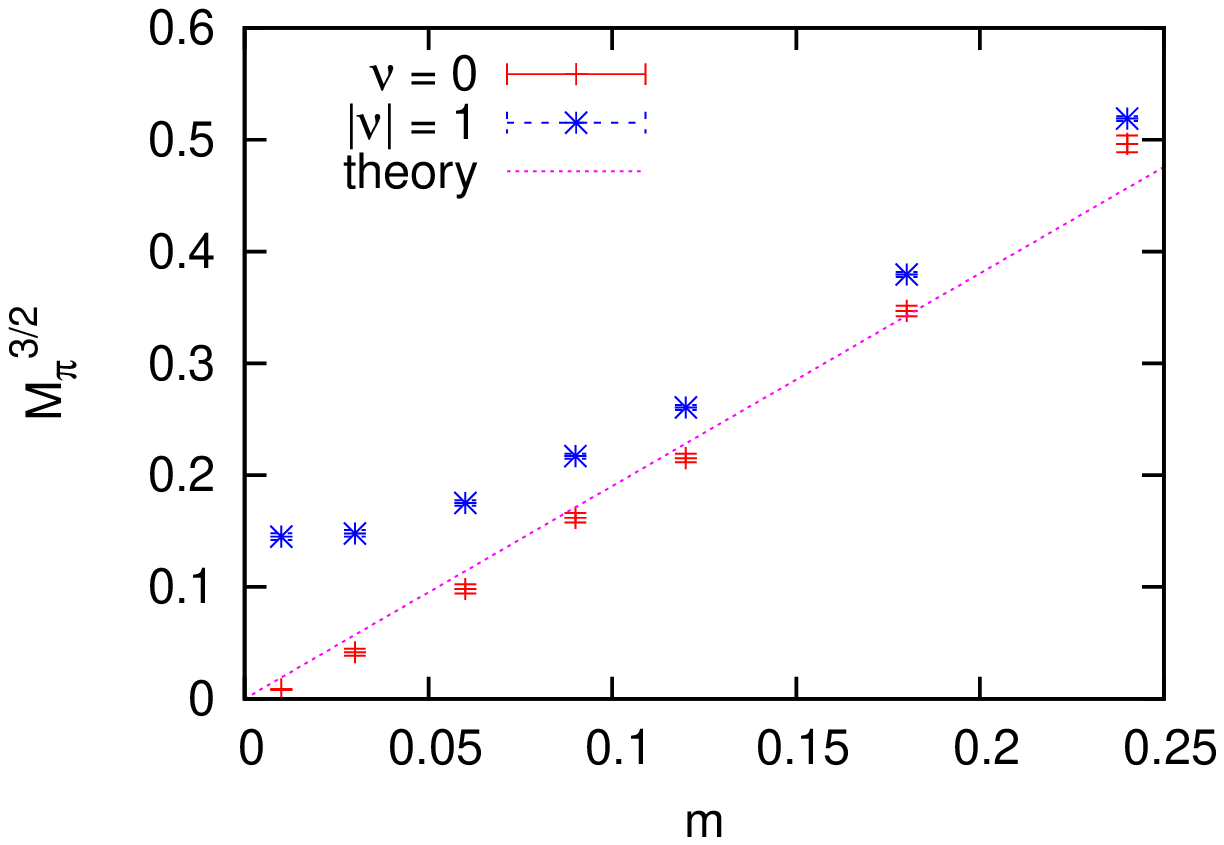}
\hspace*{-5mm}
\includegraphics[angle=270,width=.5\linewidth]{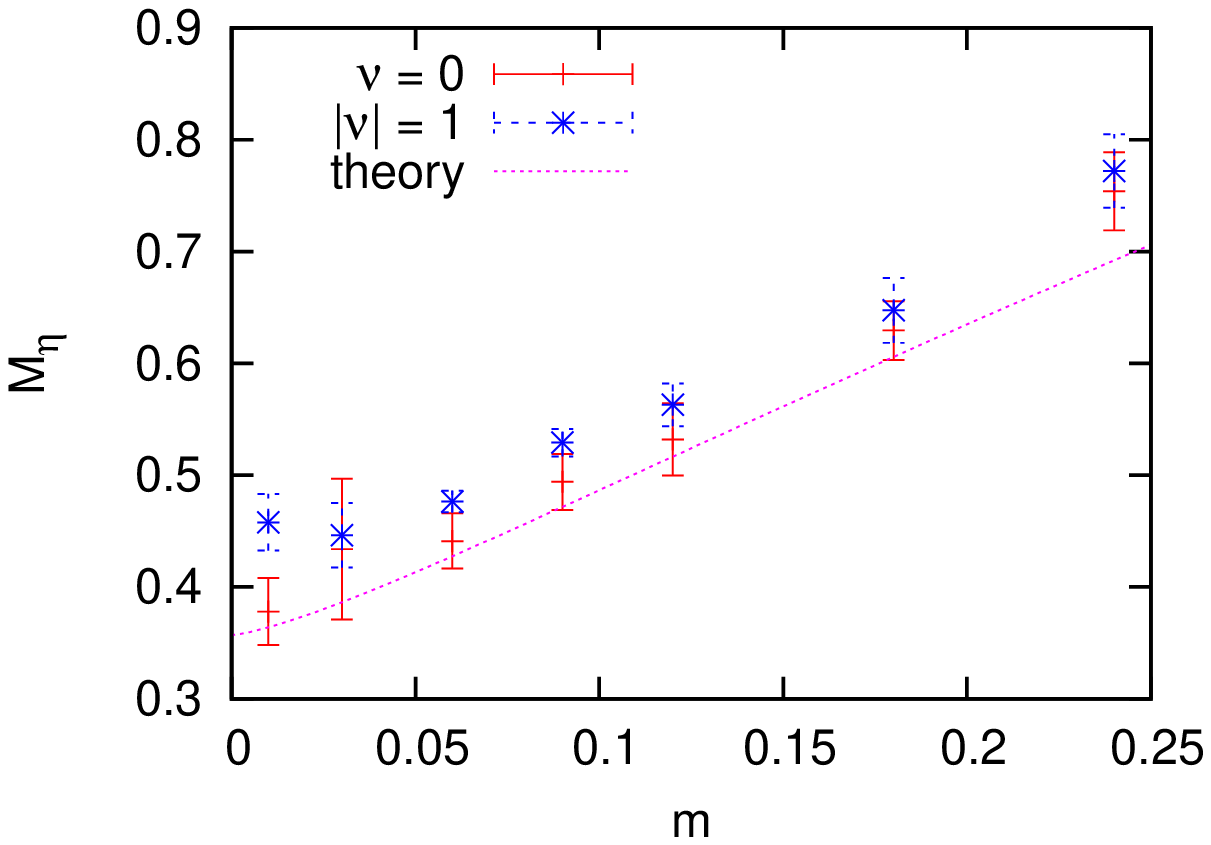}
\end{center}
\vspace{-2mm}
\caption{$M_{\pi}^{3/2}$ and $M_{\eta}$ as functions of the 
fermion mass $m$ at $L=16$, measured in the sectors $|\nu | =0$ and $1$.
For comparison we show the theoretical prediction
(summed over all topologies).}
\label{mesofig}
\vspace{-2mm}
\end{figure}

We measured the ``meson masses'' based on current correlators,
which is most efficient in this model \cite{Ivan}, and we show the
results for $L=16$ in Figure \ref{mesofig}. At least for $m=0.01$
we are clearly in the $\epsilon$-regime. 
For increasing $m$ the topological distinction shrinks.
The masses measured
in $|\nu | = 0$ and $1$ are compared to the predictions of 
Refs.\ \cite{Smilga,Ivan} (which refer to $m \ll g \simeq 0.45$).

At last we apply Method 2 to the pion mass, as it was originally
intended \cite{BCNW}. We fix $m=0.01$, $| \nu |=1$ and employ
the formula (\ref{sumfo}), 
$M_{\pi ,1} \approx M_{\pi} - A/V + B /V^{2}$.
We have results from three volumes, which we insert and solve for 
$M_{\pi}$,
\be
\{ \ M_{\pi ,1}^{(L=16)} = 0.276(4) \, , \
     M_{\pi ,1}^{(L=20)} = 0.214(4) \, , \
     M_{\pi ,1}^{(L=32)} = 0.135(4) \ \} \ \ \Rightarrow \ \
M_{\pi} = 0.078(8) \ .
\ee
This is indeed compatible with the theoretical value in
eq.\ (\ref{cond}), $M_{\pi} =0.0713 \dots$ Considering the large
$M_{\pi ,1}^{(L)}$ masses that we started from, this agreement
is an impressive success of this method.

\vspace*{-1mm}
\section{Conclusions}

Based on our experience, it appears possible --- at least in some cases 
--- to derive observables, which are properly summed over all topologies,
even if only measurements in a few fixed sectors are available.
This property (which agrees with Ref.\ \cite{Jap2})
is crucial for the future of dynamical overlap fermion
simulations, in particular in the $\epsilon$-regime, where topology
is essential \cite{LeuSmi}. The exact results and their reliability depends
on subtleties and needs to be explored further --- the Schwinger
model is ideal for such tests before large-scale QCD applications. \\

\vspace*{-2mm}
\noindent
{\bf Acknowledgements :} We thank S.\ Shcheredin and J.\ Volkholz for
their contributions to this on-going project, and S.\ D\"{u}rr for
helpful comments. The simulations were performed on the 
clusters of the ``Norddeutscher Verbund f\"ur Hoch- und 
H\"ochstleistungsrechnen'' (HLRN).

\end{document}